
\input phyzzx
\vsize 21.5 truecm
\sequentialequations

\overfullrule=0pt

\def\LamQCD{\hbox{$\Lambda_{QCD}$}}

\def\({\left\lbrack}           \def\){\right\rbrack}
\def\{{\left\lbrace}           \def\}{\right\rbrace}

\def\Fmn{F_{\mu\nu}}           \def\FMN{F^{\mu\nu}}
\def\Fls{F_{\lambda\sigma}}    
\def\Fnl{F_{\nu\lambda}}       
\def\Hmn{H_{\mu\nu}}           
\def\Hls{H_{\lambda\sigma}}    
\def\Gmn{G_{\mu\nu}}           \def\GMN{G^{\mu\nu}}
\def\Lmn{L_{\mu\nu}}           \def\LMN{L^{\mu\nu}}
\def\Rmn{R_{\mu\nu}}           \def\RMN{R^{\mu\nu}}

\def\dmu{\partial_\mu}         \def\dnu{\partial_\nu}
         
\def\dl{\partial_\lambda}         \def\ds{\partial_\sigma}
\def\Al{A_\lambda}   \def\As{A_\sigma}   
\def\Gl{G_\lambda}   \def\Gs{G_\sigma}   
   
\def\Jl{J_\lambda}   \def\Js{J_\sigma}   \def\Jr{J_\rho}

\def\Ll{L_\lambda}      
\def\Rl{R_\lambda}   \def\Rs{R_\sigma}   \def\Rr{R_\rho}
\def\slasha#1{\setbox0=\hbox{$#1$}#1
\hskip-\wd0\hbox to\wd0{\hss\sl/\/\hss}}
\def\slashb#1{\setbox0=\hbox{$#1$}#1\hskip-\wd0\dimen0=5pt\advance
       \dimen0 by-\ht0\advance\dimen0 by\dp0\lower0.5\dimen0\hbox
         to\wd0{\hss\sl/\/\hss}}

\def\g5{\gamma_5}

\def\eMNLS{\epsilon^{\mu\nu\lambda\sigma}}

\def\eMNLSR{\epsilon^{\mu\nu\lambda\sigma\rho}}
\def\eff{{\kern-0.1em{\it eff}\,}}
         
 \def\Jbm{\bar J_\mu} \def\Jbn{\bar J_\nu}
\def\Jbl{\bar J_\lambda} \def\Jbs{\bar J_\sigma}

\REF\Kaplan{D.~B.~Kaplan,
{\sl Phys.\ Lett.}{\bf B235}(1990)163;
{\sl Nucl.\ Phys.}{\bf B351}(1991)137.}

\REF\GM{A.~Manohar and H.~Georgi,
{\sl Nucl.\ Phys.} {\bf B234}(1984)189.}

\REF\Skyrme{T.~H.~R.~Skyrme, {\sl Proc.\ Roy.\ Soc.\ London}
{\bf B260}(1961)127.}

\REF\SM{E.~Witten, {\sl Nucl.\ Phys.} {\bf B223}(1983)422 and
433;
G.~Adkins, C.~Nappi, and E.~Witten,
{\sl Nucl.\ Phys.}\ B{\bf 228}(1983)552;
$SU(3)_f$ extension in E.~Guadagnini,
{\sl Nucl.\ Phys.}\ B{\bf 236}(1984)35;
P.~O.~Mazur, M.~A.~Nowak, and \hbox{M.~Prasza\l owicz},
{\sl Phys.\ Lett.}\ B{\bf 147}(1984)137.}

\REF\WittenLewes{E.~Witten in {\em Lewes Workshop Proc.;}
A.~Chodos {\em et al.}, Eds; Singapore, World Scientific, 1984.}

\REF\staticQ{G.~Gomelski, M.~Karliner and S.~B.~Selipsky,
{\sl Phys. Lett. } {\bf B323}(1994)182.}

\REF\EFHK{J.~Ellis, Y.~Frishman, A.~Hanany and M.~Karliner,
{\sl Nucl.\ Phys.}\ {\bf B382}(1992)189.}

\REF\FHK{Y. Frishman, A. Hanany and M. Karliner
{\sl Nucl. Phys.} {\bf B424}(1994)3.}

\REF\AndBon{A.~A.~Andrianov and L.~Bonora,
{\sl Nucl.\ Phys.} {\bf B233}(1984)232.}

\REF\Balog{J.~Balog,
{\sl Phys.\ Lett.}  {\bf B149}(1984)197.}

\REF\Manes{J.~L.~Manes,
{\sl Nucl.\ Phys.} {\bf B250}(1985)369.}

\REF\BarZu{W.~A.~Bardeen and B.~Zumino,
{\sl Nucl.\ Phys.} {\bf B244}(1984)421.}

\REF\DS{
P.H. Damgaard and R. Sollacher,
{\sl Phys. Lett.} {\bf B322}(1994)131.}

\REF\Aitchison{
I. Aitchison,
C. Fraser, E. Tudor and J. Zuk,
{\sl Phys. Lett.} {\bf 165B}(1985)162.}

\nopubblock
\baselineskip=15pt
\line{\hfill WIS-95/25/June-PH}
\line{\hfill TAUP-2264-95}
\line{\hfill hep-ph/9507206}
\line{\hfill}
\title{On the Stability of Quark Solitons in
 QCD\foot{Research supported by the Israel Science Foundation
   administered by the Israel Academy of Sciences and Humanities.}}
\author {
Yitzhak Frishman\foot{e-mail: fnfrishm@wicc.weizmann.ac.il}
and Amihay Hanany\foot{e-mail: ftami@wicc.weizmann.ac.il}
}
\address{ Department of Particle Physics   \break
          Weizmann Institute of Science \break
          76100 Rehovot Israel}
\author {Marek Karliner\foot{e-mail: marek@vm.tau.ac.il}}
\address{School of Physics and Astronomy\break
         Beverly and Raymond Sackler    \break
         Faculty of Exact Sciences      \break
         Ramat Aviv Tel-Aviv, 69987, Israel}
\abstract{
\tenpoint\baselineskip=11pt
We critically re-examine our earlier
derivation of the effective low energy action for QCD in 4
dimensions with chiral fields
transforming non-trivially under both color and flavor,
using
the method of anomaly integration.
We find several changes with respect to our previous
results, leading to much more compact expressions, and making
it easier to compare with results of other approaches to the same
problem.
With the amended effective action, we find that
there are no stable soliton solutions.
In the context of the quark soliton program,
we interpret this as an indication that the full low-energy
effective action
must include additional terms,
reflecting
possible modifications at short distances and/or
the non-trivial structure
of the gauge fields in the vacuum,
such
as $\VEV{F_{\mu\nu}^2}\neq 0$.
Such terms are absent in the formalism based on anomaly
integration.}
\endpage

\doublespace
\baselineskip=25pt
\chapter{Introduction and Motivation}
In the recent years some progress have been made towards
establishing the connection between
the phen\-o\-men\-ologically successful
non-relativistic con\-stit\-uent-quark model (NRQM),
and QCD's fundamental degrees of freedom.
Kaplan\refmark{\Kaplan} proposed a physical picture
combining  some  of the  features  of  the
chiral  quark model\refmark{\GM}
and the
skyrmion\refmark{\Skyrme-\WittenLewes}.
It was postulated that at
distances smaller than the confinement scale but large enough to allow
for nonperturbative phenomena the effective dynamics of QCD is
described by chiral dynamics of a bosonic field which takes values in
$U(N_c {\times} N_f)$.  This effective theory admits classical
soliton solutions. Assuming that they are stable and may be quantized
semiclassically, one then finds that these solitons are extended
objects  with  spin $1/2$,  and that they  belong  to the  fundamental
representation of color and flavor.  Their mass is of order \LamQCD\
and radius of order $1/\LamQCD$.\refmark{\staticQ}
 It is very tempting to identify them
as the constituent quarks. Thus the constituent quarks in this model
are ``skyrmions'' in color space.

It turns out that in 2 space-time dimensions this picture
is exact\refmark{\EFHK}. Thanks to exact non-abelian bosonization
one can rewrite the action of $QCD_2$ in terms of purely bosonic
variables, which are chiral fields $\in U(N_c{\times}N_f)$.
It is then straightforward to demonstrate that the
only non-trivial static solutions of the classical equations of
motion are those which contain either a soliton and an anti-soliton
or $N_c$ solitons. The solitons transform under both flavor and color,
yet their bound states are color singlets and have the quantum
numbers of baryons and mesons.

In four space-time dimensions there is no exact bosonization,
and therefore any attempt at derivation of a similar picture
in four dimensions must rely on certain approximations.
In our previous work\refmark{\FHK}
 we have derived the approximate low-energy
effective chiral lagrangian with target space in
$ U(N_c{\times}N_f)$ using the approach based on integration of
the anomaly equations\refmark{\AndBon-\BarZu}.
Equivalent results were independently obtained
in ref.~\DS, using an a priori different approach.

The purpose of the current work is to critically re-examine
the results of
ref.~\FHK, with particular emphasis
on the question whether the action we have derived
can support stable, time independent classical solutions.
The existence of such solutions appears to us to be a necessary
condition for establishing the physical picture in which
constituent quarks are solitons of a low-energy effective action
of QCD in four dimensions.

We find several changes with respect to our previous
results, leading to much more compact expressions, and making
it easier to compare with results of other approaches to the same
problem, and in particular with the action proposed by
Kaplan\refmark{\Kaplan}.

With the amended effective action, we find that
there are no stable soliton solutions.
In the context of the quark soliton program,
we interpret this as an indication that the full low-energy
effective action
must include additional terms,
reflecting
possible modifications at short distances and/or
the non-trivial structure
of the gauge fields in the vacuum,
such
as $\VEV{F_{\mu\nu}^2}\neq 0$.
Such terms are absent in the formalism based on anomaly
integration.

The layout of the paper is as follows. In Section 2 we
rederive
the low energy classical action, resulting in much more compact
final expressions.
In Section 3 we examine the necessary conditions for existence
of stable, time independent classical solutions with finite energy,
and
conclude that in order for such solutions to exist, the
effective action must contain additional terms which are not present
in our derivation.
Section 3 is devoted to discussion and interpretation of the results.

\chapter{Derivation of the effective action}
We follow the conventions and notation of ref.~\FHK.

The variation of the determinant
under the axial transformation of the external fields,
including terms up to zeroth power
of the cutoff $\Lambda$, is given by
$$\eqalign{-i&{\delta\log Z\over\delta\lambda}={1\over4\pi^2}
\biggl\lbrace\eMNLS\biggl[
{1\over4}\Fmn\Fls + {1\over12} \Hmn\Hls - {2i\over3}
(A_\mu A_\nu\Fls+A_\mu\Fnl\As  \cr
&+\Fmn \Al\As) - {8\over3} A_\mu A_\nu\Al\As\biggr]
+16\Lambda^2 D_\mu A^\mu \cr
&+{2i\over3}\(D^\mu\Fmn,A^\nu\)+{i\over3}\(\Fmn,D^\mu A^\nu\) \cr
&+{1\over3}\{D^\mu A_\mu,A_\nu A^\nu\}-2A_\mu D_\nu A^\nu A^\mu-  \cr
&-{2\over3}\{D^\mu A^\nu,\{A_\mu, A_\nu\}\}
+ {1\over3} (D^\mu D_\mu D^\nu A_\nu)\biggr\rbrace
+{\cal O} (\Lambda^{-2}) \cr
}\eqn\Anofd$$
where $\Fmn=\dmu V_\nu-\dnu V_\mu+i\(V_\mu,V_\nu\)+i\(
A_\mu,A_\nu\)={1\over2}(\Lmn+\Rmn),$
$D_\mu A_\nu = \partial_\mu A_\nu + i \(V_\mu,A_\nu\)$
and $\Hmn={1\over2}(\Rmn-\Lmn)=(D_\mu A_\nu - D_\nu A_\mu)$.
(cf. eq. (4.1) of ref.~\FHK). $V_\mu$ is the external source coupled to the
vector current of the fermions, while $A_\mu$ is the external source coupled to
the axial current of the fermions.

An action which has equation \Anofd\ as its variation is given by
$$ \eqalign{-i\log Z_1=&-{2\Lambda^2\over\pi^2}\int\Tr(A_\mu A^\mu)
+S_{CS}^5\(R\)-S_{CS}^5\(L\)\cr
&-{i\over48\pi^2}\int\eMNLS\Tr\({i\over 2}(\Rmn+\Lmn)\{\Ll,\Rs\}
+(L_\mu L_\nu-{1\over2}L_\mu R_\nu+R_\mu R_\nu)\Ll\Rs\)\cr
&-{1\over12\pi^2}\int\Tr\{{1\over4}(\Fmn)^2-i\FMN\(A_\mu,A_\nu\)
-{1\over2}(D_\mu A^\mu)^2-(A_\mu A_\nu)^2\}}\eqn\Acanfd $$

where $S_{CS}^5$ is the five dimensional Chern-Simons action
\nextline
\vbox{
$$S_{CS}^5\(R\)={1\over24\pi^2}\int\eMNLSR\Tr(R_\mu\dnu\Rl\ds\Rr+
{3\over2}iR_\mu R_\nu\Rl\ds\Rr-{3\over5}R_\mu R_\nu\Rl\Rs\Rr).\eqn\CSf$$
To derive equation \Acanfd, it is useful to employ
$$\eqalign{
\delta S_{CS}^5\(R\)&={1\over8\pi^2}\int d^5x\eMNLSR\Tr\(\delta R_\mu\left(
\dnu\Rl\ds\Rr+i\{R_\nu\Rl,\ds\Rr\}-R_\nu\Rl\Rs\Rr\right)\)\cr &+
{1\over24\pi^2}\int d^4x\eMNLS\Tr\(\delta R_\mu\left(\{R_\nu,\dl\Rs\}
+{3\over2}iR_\nu\Rl\Rs\right)\).}\eqn\CSf$$
For the special case where
$\delta R_\mu=\dmu\delta\omega+i\(R_\mu,\delta\omega\)$ we have
$$\delta S_{CS}^5\(R\)={1\over24\pi^2}\int d^4x\eMNLS\Tr\(\delta\omega
\left(\dmu R_\nu\dl\Rs
+{1\over2}i\{R_\mu R_\nu,\dl\Rs\}-{1\over2}iR_\mu\dnu\Rl\Rs\right)\).\eqn\CSf$$
The resulting effective action takes the form
$$\eqalign{S_\eff&={\Lambda^2\over{2\pi^2}}\int\Tr\(D_\mu U D^\mu U^{-1}
-(L_\mu - R_\mu)^2\)\cr
&-{1\over240\pi^2}\int\eMNLSR\Tr(J_\mu J_\nu\Jl\Js\Jr)\cr
&+{i\over48\pi^2}\int\eMNLS\Tr\biggl\lbrace {i\over2}\Rmn\{\Rl,\Jbs\}+R_\mu
R_\nu\Rl\Jbs\cr
& + {1\over2}R_\mu\Jbn\Rl\Jbs-\Jbm\Jbn\Jbl\Rs\biggr\rbrace\cr
&-{i\over48\pi^2}\int\eMNLS\Tr\{{i\over2}(\Rmn+\Lmn)\{\Ll,\Rs\}
+(L_\mu L_\nu-{1\over2}L_\mu R_\nu+R_\mu R_\nu)\Ll\Rs\}\cr
&+{i\over48\pi^2}\int\eMNLS\Tr\{{i\over2}(U^{-1}\Rmn U+\Lmn)\{\Ll,\Rs^U\}
+(L_\mu L_\nu-{1\over2}L_\mu R^U_\nu+R^U_\mu R^U_\nu)\Ll\Rs^U\}\cr
&+{1\over192\pi^2}\int\Tr\biggl\lbrace[
2(D_\mu (U^{-1} D^\mu U))^2-(D_\mu U^{-1}
D_\nu U)^2\cr
&-{4i}(\Lmn D^\mu U^{-1}D^\nu U+\Rmn D^\mu UD^\nu U^{-1})+
2(U^{-1}\Rmn U\LMN-\Lmn\RMN)]\cr
&+{16} [i\FMN\(A_\mu,A_\nu\)+{1\over2}(D_\mu A^\mu)^2+
(A_\mu A_\nu)^2)]  \biggr\rbrace
}\eqn\ealr$$
}
where $D_\mu U=\dmu U+iR_\mu U-iUL_\mu$,
$D_\mu U^{-1}=\dmu U^{-1}+iL_\mu U^{-1}-iU^{-1}R_\mu$,
$J_\mu=U^{-1}i\dmu U$ and $\Jbm=-UJ_\mu U^{-1}$.

In order to make contact with the usual
formulation of QCD,  we write down the
effective action for
$L_\mu=R_\mu=G_\mu$ in eq. \ealr, to obtain
$$\eqalign{S_\eff&={\Lambda^2\over2\pi^2}\int\Tr(D_\mu UD^\mu U^{-1})
-{1\over240\pi^2}\int\eMNLSR\Tr(J_\mu J_\nu\Jl\Js\Jr)\cr
&+{i\over48\pi^2}\int\eMNLS\Tr\{{i\over2}\Gmn\{\Gl,\Jbs\}+(G_\mu G_\nu
+{1\over2}G_\mu\Jbn
+\Jbm\Jbn)\Gl\Jbs
\}\cr
+{i\over48\pi^2}\int\eMNLS&\Tr\({i\over2}(U^{-1}\Gmn U+\Gmn)\{\Gl,\Gs^U\}
+(G_\mu G_\nu-{1\over2}G_\mu G^U_\nu+G^U_\mu G^U_\nu)\Gl\Gs^U\)\cr
&+{1\over192\pi^2}\int\Tr\{2\(D_\mu(U^{-1}D^\mu U)\)^2-(D_\mu U^{-1}D_\nu
U)^2\right.\cr
&\left.-4i\Gmn\(D^\mu U^{-1},D^\nu U\)+2(U^{-1}\Gmn
U\GMN-\Gmn\GMN)\},\cr}\eqn\eag$$
where $G^U_\mu =U^{-1}G_\mu U-U^{-1}i\dmu U
=U^{-1}(G_\mu+\Jbm)U$.

\chapter{Stability analysis of classical solutions}

Our initial hope was to find nontrivial stable
minima of the action \eag. In the process of looking for
such solutions, we observed some numerical instabilities,
which caused us to suspect that the action \eag\ might
be unbounded from below.

In order to demonstrate that this
is indeed the case, it is sufficient to show this for
one trial set of functions, $U_{trial}(x)$ and $G_{trial}(x)$.
In order to simplify the stability analysis,
we will therefore begin with
classical action without gauge
fields, $G_{trial}(x)=0$.

Had such an action lead to stable soliton solutions,
it would amount to an (approximate) bosonization of free quarks
in four dimensions, which would have been an interesting result
in its own right.
As the following shows, this has not
been attained in the present formalism. We shall comment later
on what we believe might be the
possible reasons for this.

We therefore set the gauge field to zero, resulting  in
$$\eqalign{S_\eff\(0,0,U\)&={\Lambda^2\over2\pi^2}\int\Tr(J_\mu J^\mu)
-{1\over240\pi^2}\int\eMNLSR\Tr(J_\mu J_\nu\Jl\Js\Jr)\cr
&-{1\over192\pi^2}\int\Tr\lbrack
                            2(\dmu J^\mu)^2+(J_\mu J_\nu)^2 \rbrack
                             \cr}\eqn\Sezes$$

Motivated by the Skyrme model,
we are looking for a radially symmetric hedgehog solution.
We choose the classical solution to be a field of the form
$$U_c=e^{if(r)\vec\tau\cdot\hat r}$$
where $\vec\tau$ are pauli matrices, the generators of some $SU(2)$ subgroup
of $U(N_cN_f)$ and $f(r)$ is a radial shape function with boundary
conditions $f(0)=\pi$, $f(\infty)=0$.
The choice of the embedding will become relevant only if
stable solutions exists, and then it should be discussed
together with quantization of the collective coordinates.
The Wess-Zumino term vanishes and the rest of the terms
are given by
$$-(J_\mu^c)^2=(f')^2+{2\sin^2f\over r^2} $$
$$(J_\mu^cJ_\nu^c)^2=(f')^4-{4(f')^2\sin^2f\over r^2}
\eqn\JJterms$$
$$(\dmu J^{c\mu})^2={ \left(    f''   +{2   f'\over r}
            -{ \sin^2f\over r^2}\right) }^2           $$

Next,
we define an effective potential $V_{\eff}$ as
minus the action divided by $N_cN_f$ and integrated over space only,

$$\eqalign{V_\eff(f)&={2\Lambda^2 \over\pi}\int_{0}^{\infty}dr\,\lbrack{r^2
 (f')^2+ 2 \sin^2f}\rbrack \cr &+{1\over{48 \pi}} \int_{0}^{\infty}dr \left[
 2{\left( r f''+ 2f' -{\sin(2f)\over r}\right)}^2+\lbrack r^2
 (f')^4-{4(f')^2\sin^2f} \rbrack \right]\cr}\eqn\V$$

Finding stable solutions of the action \Sezes\ is now reduced to
functional minimization of the effective potential \V.

Consider a family of trial functions
$$f(r)=2\tan^{-1}({a\over r})\eqn\trialFamily$$
where $a$ is a variational parameter
which also determines the soliton size.
The functions \trialFamily\
satisfy the boundary conditions at both $r=0$ and $r\rightarrow
\infty$:

$$\eqalign{
f(r=0) &= \pi\crr
f(r\rightarrow \infty) &=0
}\eqn\boundaryconds$$

The first condition is needed to ensure that the solution carries
one unit of winding number, which
in our normalization corresponds to one quark.   \par

For this family of trial functions,
$$V_\eff (a)=6{\Lambda^2}a -{ 1\over 96a}\eqn\Veffofa$$
Hence for $a\rightarrow 0$
the potential is unbounded below. Thus, when attempting to solve
the equations of motion, we will find that the soliton
profile will be ``squeezed'' to zero width around the origin.
This shrinking of the classical solution to zero size
occurs despite the presence of both two- and four-derivative
terms in the action \Sezes.

A similar effect occurs already in the $\sigma$
model\refmark{\Aitchison}. Also there, the approximate action
with up to four derivatives on the $\sigma$ and $\vec \pi$
fields, constrained to $\sigma^2 + {\vec\pi}^2=f^2$ has its
classical solutions ``squeezed" to zero width, and
with energy tending to \ $-\infty$.

In two dimensions, stabilization of the soliton is provided by
the mass term. In four dimensions, one may add a mass term
$$2 m_{Q}^4 \left[ 1-\cos(f) \right]\eqn\massTerm$$
where $m_{Q}$ is some scale related to the original quark mass
in the QCD Lagrangian,
(not quite the bare mass itself,
as there is normal ordering to be performed;
see Appendix of ref.~[\EFHK]). Such a mass term does not provide
the desired stabilization, however.
For the trial function above, this term will have a divergent contribution
to the potential coming from the integration over
large $r$. Since the stabilization problem occurs at small distances,
this large-$r$ divergence due to the mass term cannot cure the problem.
In order to isolate the large-$r$ divergence,
we will treat the problem
of large $r$ by putting a cut-off $R$.
We expect that eventually
such a cut-off will actually be provided
by the confinement in QCD. \par
Now the contribution of the mass term
to the potential will be
$$16 \pi m_{Q}^4 a^3 \left[ {R\over a}-\tan^{-1}({R\over a})\right]
\eqn\massConToV
$$
which tends to zero as ``a'' tends to zero, thus not changing the fact that
the potential is unbounded below for small scales $a$.

\chapter{Discussion and Interpretations}

There are various options to overcome these difficulties.
The first is by choosing a different regularization scheme.
This may change the coefficients of the dimension four operators which appear
in the Lagrangian.
In particular it may change the terms in such a way that we will have
a commutator squared as in the Skyrme model. This probably is the
only known action which produces a positive Hamiltonian. A second way
out of this problem is to refer to non-perturbative corrections which
will change the form of the coefficients in such a way as to get a
Skyrme like action.

We should remark, however, that in general we would not expect
a scheme change to influence physical results, like the emergence
of constituent quarks. It may happen, however, that
due to the approximations made, we may be able to derive certain
quantities in one scheme and not in another.

Recall that in two dimensions, the scheme was completely
fixed by requiring vector conservation and that the axial
be the dual of the vector. The latter requirement was a
result of our wish to have the bosonic version correspond to
the fermionic one, and in the latter the axial is indeed the
dual of the vector (see our work, ref. [\FHK] for details).
We do not have an analogous requirement in four dimensions
as yet.

Let us also remark that our classical configurations tend to be
``squeezed" to zero size, and with energy tending to
\ $-\infty$. The troublesome part is at short distances. But this
is precisely the regime of high momenta, where our approximations
are inadequate, as we have neglected terms with six derivatives
or more.
So we either have to find a better approximation, or maybe
exclude some short distance region.

A final comment. We expect
the effective action \eag, after integrating out the
gauge fields and taking trace over color, to yield
an effective action in flavor space. But due to the
non-positive nature of the potential that we discovered above,
we do not expect, within the present approximation, to
get the Skyrme model with the assumed standard
positive-definite stabilizing four-derivative
term\refmark{\Skyrme,\SM}.

\bigskip
\vbox{
\ack
This research was supported by the Israel Science Foundation
   administered by the Israel Academy of Sciences and Humanities.
The research of M.K.
was supported in part by
the Weizmann Center at the Weizmann Institute and by
a Grant from the G.I.F., the
German-Israeli Foundation for Scientific Research and
Development.
Y.F. would like to thank I.~Klebanov,
for pointing out the existence of ref.~\Aitchison.
}

\bigskip
\refout
\end